\documentclass[aps,prl,twocolumn,amsmath,amssymb,showpacs]{revtex4-1}

\usepackage{mathtools}
\usepackage{amsmath}
\usepackage{float}
\usepackage{graphicx,array}
\usepackage{color}
\usepackage{epsfig}
\usepackage{latexsym,cases,soul}
\usepackage{bm}

\usepackage[colorlinks=true, allcolors=blue]{hyperref}
\usepackage{cleveref}
\usepackage{color}
\usepackage[normalem]{ulem}

\newcommand{\be}{\begin{equation}}
\newcommand{\ee}{\end{equation}}
\newcommand{\bea}{\begin{eqnarray}}
\newcommand{\eea}{\end{eqnarray}}

\def\bsn{\begin{subnumcases}}
\def\esn{\end{subnumcases}}

\def\KJ#1{\textcolor{black}{#1}}

\def\rev#1{\textcolor{black}{#1}}

\def\newrev#1{\textcolor{black}{#1}}

\begin{document}

\title{Current fluctuations in symmetric zero-range process below and at critical density}

\author{Tanmoy Chakraborty\textsuperscript{$\star$}, Punyabrata Pradhan\textsuperscript{$\star$} and Kavita Jain\textsuperscript{$\dagger$}} 

\affiliation{${}^\star$ Department of Physics of Complex Systems, S. N. Bose
   National Centre for Basic Sciences, Block-JD, Sector-III, Salt Lake,
   Kolkata 700106, India
\\
${}^\dagger$ Theoretical Sciences Unit,
Jawaharlal Nehru Centre for Advanced Scientific Research,
Bangalore 560064, India}


\date{\today}

\begin{abstract}
Characterizing current fluctuations in a steady state is of fundamental interest and has attracted considerable attention in the recent past. However, the bulk of the studies are limited to systems that either do not exhibit a phase transition or are far from criticality. Here we consider a symmetric zero-range process on a ring that is known to show a phase transition in the steady state. 
\rev{We analytically calculate two density-dependent transport coefficients, namely, the bulk-diffusion coefficient and the particle mobility,} that characterize the first two cumulants of the time-integrated current. We show that on the hydrodynamic scale, away from the critical point, the variance of the time-integrated current in the steady state grows with time $t$ as $\sqrt{t}$ and $t$ at short and long times, respectively.  Moreover, we find an expression of the full scaling function for the variance of the time-integrated current and thereby the amplitude of the temporal growth of the current fluctuations.  
At the critical point, using a scaling theory, we find that, while the above-mentioned long-time scaling of the variance of the cumulative current continues to hold, the short-time behavior is anomalous in that the growth exponent is larger than one-half and varies continuously with the model parameters.   
\end{abstract}

\maketitle


\noindent{\bf Introduction:} Unlike in thermodynamics where one deals with averages, at the microscopic level, the density and current are in fact stochastic. Indeed, characterization of their fluctuations is a fundamental problem in statistical physics and has attracted considerable interest.  
The behavior of current distribution and its cumulants have been investigated in a variety of classical models including simple exclusion process \cite{Derrida:1998, Bodineau:2004, Derrida:2007, Appert-Rolland:2008, Derrida:2009, Imparato:2009, Krapivsky:2012}, contact process \cite{Gorissen:2009}, zero-range process \cite{Harris:2005}, models of heat conduction \cite{Kipnis:1982, Basile:2006, Hurtado:2009}, multi-species chemical reactions \cite{Fiore:2021}, sandpiles \cite{Mukherjee_PRE2023, Mukherjee_condmat2024}, and more recently, active matter \cite{GrandPre:2018, Banerjee:2020, Agranov:2023, Jose:2023, Chakraborty:2024}, using microscopic theory, Monte-Carlo simulations or in the hydrodynamic framework of macroscopic fluctuation theory \cite{Bertini:2001, Bertini:2002, Bertini:2015, Derrida:2007}. 

These studies have shown that on the hydrodynamic scale, the steady-state current statistics for finite systems depend on whether the time is smaller or larger than the (system size-dependent) relaxation time scale. 
 Specifically, in a one-dimensional diffusive system of size $L$, if $Q_i(t)$ is the time-integrated current across a bond $(i,i+1)$ in time interval $t \gg L^2$  with density $\rho$ and $\rho+\Delta \rho$ at the left and right boundary, the average current typically behaves as \cite{Bertini:2001, Bertini:2002}
 \be
\lim_{t \to \infty} \frac{\langle Q_i(t) \rangle}{t} = -D(\rho) \frac{\Delta \rho}{L}, \label{mean1}
\ee
which is the well-known Fick's law; here $D(\rho)$ is the bulk-diffusion coefficient that characterizes the response to a density gradient. 
For $\Delta \rho=0$ (or, on a periodic domain), although the average current is zero, the second cumulant (that is, the variance) of time-integrated current scales as  \cite{Bertini:2001, Bertini:2002}
\be 
\lim_{t \to \infty} \frac{\langle Q^2_i(t) \rangle_c}{t} = \frac{2\chi(\rho)}{L}, \label{var1}
\ee
where $\chi(\rho)$ is the collective particle mobility.  
However, in an infinitely large system (or, equivalently, in the time domain $t \ll L^2$ for a finite system), the above scaling does not hold, and the variance of the time-integrated current grows the same way as the variance of the individual tagged-particle displacement in single-file transport \cite{Alexander:1978}, namely, as $ \sqrt{t}$ \cite{DeMasi:2002,Derrida:2009, Barkai:2009,Krapivsky:2014,Sadhu:2016}. \rev{The root-mean-squared displacement of a tagged particle can also be mapped to height fluctuations in one-dimensional growing interfaces whose scaling belongs to the Edwards-Wilkinson or the Kardar-Parisi-Zhang universality class for diffusive or driven system, respectively \cite{Majumdar:1991, Kriecherbauer:2010}.}

However, in a system undergoing a phase transition, the behavior of current fluctuations can change at the critical point. For example, in the 
ABC model \cite{Evans:1998} that shows phase separation in the stationary state; at the critical point, all the time-integrated current cumulants scale linearly with time, but the coefficient of the growth law displays anomalous scaling with the system size \cite{Gerschenfeld:2011}. On the other hand, in a totally asymmetric exclusion process with open boundaries, the scaling of the current fluctuation at the transition line are numerically found to remain the same as in the phases separated by it  \cite{Gorissen:2009}. 
Recently, in the context of absorbing-phase transition, in the thermodynamic limit, the time-integrated bond-current variance in the one-dimensional conserved-mass stochastic sandpiles near criticality has also been found to grow anomalously with a growth exponent smaller than one-half \cite{Mukherjee_PRE2023, Mukherjee_condmat2024}. 

Evidently, the current statistics at the critical point can be different from those away from it, yet they remain largely unexplored. 
In this Letter, we study steady-state current fluctuations in a zero-range process (ZRP) -- a paradigmatic model of an interacting particle system that, in a certain parameter regime, shows a phase transition between a fluid phase and a condensate phase in which fluid and a macroscopic condensate coexist \cite{Evans:2005}. 
We find that,  in regions of the parameter space where the bulk-diffusion coefficient is finite, the sub-diffusive and linear scaling of current fluctuations mentioned in the Introduction hold. But, at the critical point, for a range of model parameters, the bulk-diffusion coefficient vanishes in the thermodynamic limit, and as a consequence, the variance of the time-integrated current increases algebraically in time with a continuously varying exponent larger than one-half in this regime.  In the fluid phase, these results are obtained analytically, and, at the critical point, we provide a scaling theory; the details of the calculations are provided in the Supplemental Material (SM). All the results are verified using extensive Monte Carlo simulations in which the \rev{data were obtained from over $10^4$ independent initial conditions chosen from the stationary state.}

\noindent{\bf Model and its stationary state:} We consider a ZRP on a ring with $L$ sites, each of which contains $m_i \geq 0$ particles of mass unity.  In continuous time, \KJ{a particle hops out of the $i$th site having $m_i > 0$ particles} with rate $u(m_i)$, symmetrically to \KJ{either left or right nearest neighbor}; these dynamics keep the total number of particles and global density constant at $\sum_{i=1}^L m_i=M$ and $\overline{\rho}=M/L$, respectively. In this work, the hop rate is chosen to be
\begin{eqnarray}\label{hopping_rate}
u(m)=1+\frac{b}{m}~,~b > 2.
\end{eqnarray}
\KJ{The joint mass distribution of the configurations that respect the particle number conservation is exactly given by \cite{Evans:2005}
\be
P(\{m_1,...,m_L\}) \propto \prod_{i=1}^L f(m_i) \label{prod},
\ee
where $f(m)=\frac{m!}{(b+1)_m}$ for the hop rate in Eq.~\eqref{hopping_rate}, and $(a)_m=a (a+1)... (a+m-1)$. From Eq.~\eqref{prod}, the single site particle distribution in the grand canonical ensemble can be written as
\be
p(m)=\frac{z^m f(m)}{g(z)}, \label{singlesite}
\ee 
where $0 < z(\bar{\rho})=\langle u \rangle \leq 1$ is the fugacity and $g(z)=\sum_{m=0}^\infty z^m f(m)$ is the normalization constant.  As the global density is conserved, from Eq.~\eqref{singlesite}, we have  
\be
{\bar \rho}=\frac{z}{g(z)} \frac{d g(z)}{d z}=\frac{z \, _2F_1(2,2;b+2;z)}{(b+1) \, _2F_1(1,1;b+1;z)} \label{fugacity}
\ee
where $\, _2F_1(\alpha,\beta;\gamma;z)$ is the Gauss hypergeometric function.}

Equation \eqref{fugacity} shows that the fugacity increases with total density ${\bar \rho}$, but it reaches its maximum value, {\it viz.}, one at a finite critical density \cite{Evans:2005}, 
\be
{\bar \rho}_c=\frac{1}{b-2} \label{critden}
\ee
For ${\bar \rho} < {\bar \rho}_c$, the fugacity is below one and the mass distribution $p(m)$ decays exponentially, and, at the critical density, it decays algebraically as $m^{-b}$. But for ${\bar \rho} > {\bar \rho}_c$ where $z=1$, the mean density at $L-1$ sites remains at the critical density ${\bar \rho}_c$ and the excess density, ${\bar \rho}-{\bar \rho}_c$ condenses into a single mass cluster at one site. 

\noindent{\bf Theory of bond-current fluctuations: }We are interested in understanding the statistics of the time-integrated current $Q_i(t)$ defined as the net current across the bond $(i, i+1)$ up to time $t$ in the {steady state}. 
We first consider the average bond current,  $\langle Q_i(t) \rangle$. To derive the time-evolution equation for relevant observables from microscopic dynamical rules, we note that, in the infinitesimal time interval $[t, t+dt]$, the cumulative current will change if a particle either hops in (out) site $i+1$ from (to) $i$ so that
\begin{eqnarray}\label{Q_update_eqn}
Q_{i}(t+dt) = 
\left\{
\begin{array}{ll}
Q_{i}(t) + 1 \vspace{0.25 cm}, ~~ & {\rm prob.}~~ \frac{1}{2}	{u}_i dt, \\
Q_{i}(t) - 1 \vspace{0.25 cm}, ~~ & {\rm prob.}~~ \frac{1}{2} {u}_{i+1} dt , \\
Q_{i}(t) \vspace{0.25 cm}, ~~ & {\rm prob.}~~ 1- \frac{1}{2}({u}_{i} + {u}_{i+1}) dt, \\
\end{array}
\right.
\end{eqnarray}
where, for brevity, we have defined ${u}_{i} \equiv u[{m}_i(t)]$. Using the above equation, we then obtain
\begin{eqnarray}
\label{Q_time-evolution}
\frac{d}{dt}\langle Q_{i}(t) \rangle =\frac{1}{2}\left( \langle u_{i} \rangle - \langle u_{i+1} \rangle \right) = \langle J_{i}(t)\rangle.
\end{eqnarray} 
At large times, assuming the observable $\langle u_{i} \rangle$ is governed by the corresponding slow variable, which is the local density $\rho_{i}$ [that is, $\langle u_{i} \rangle \equiv z(\rho_{i})  $], and performing a Taylor series expansion about the global density $\overline{\rho}$ on the RHS of Eq.~\eqref{Q_time-evolution}, we recover the Fick's law for the average current,
\begin{eqnarray}
   \langle J_{i} \rangle \approx D(\overline{\rho})(\rho_{i} - \rho_{i+1}), \label{currrho}
\end{eqnarray}
where the density-dependent bulk-diffusion coefficient  is given by,
\begin{eqnarray}\label{bulk-diffusivity}
 D(\overline{\rho})&=& \frac{1}{2} \left[ \frac{\partial z(\rho)}{\partial \rho} \right]_{\rho=\overline{\rho}}.
 \end{eqnarray}
As shown in Sec.~I of the SM, $D(\overline{\rho})$ decreases monotonically with the global density. For later reference,  it is important to note that the bulk-diffusion coefficient remains finite for densities below the critical density. 
But, it vanishes at the critical point for $2 < b < 3$ in the thermodynamic limit \rev{[see Eq.~\eqref{D_L_scaling_1}]}. 

 
 We now derive the time-evolution equation for the variance $\langle Q^{2}_i(t) \rangle_c$ of the time-integrated bond-current.  In a manner similar to above, we write all possible ways in which the random variable $Q^{2}_i(t)$ changes in an infinitesimal time interval $[t, t+dt]$:
\begin{eqnarray}\label{Q_sq_update_eqn}
Q^{2}_{i}(t+dt) = 
\left\{
\begin{array}{ll}
(Q_{i}(t) + 1)^{2} \vspace{0.25 cm}, ~~ & {\rm prob.}~~ \frac{1}{2}	{u}_{i} dt, \\
(Q_{i}(t) - 1)^{2} \vspace{0.25 cm}, ~~ & {\rm prob.}~~ \frac{1}{2}{u}_{i+1} dt , \\
Q^{2}_{i}(t) \vspace{0.25 cm}, ~~ & {\rm prob.}~~ 1- \frac{1}{2}({u}_{i} + {u}_{i+1}) dt. \\
\end{array}
\right.
\end{eqnarray}
Using the above rules and Eq.~\eqref{Q_time-evolution}, we find that
\begin{eqnarray}\label{Q_sq_time-evolution}
\frac{d}{dt}\langle Q^{2}_{i}(t) \rangle_c &=& \frac{1}{2}\left( \langle u_{i} \rangle + \langle u_{i+1} \rangle \right) \nonumber \\
&+& \left[ \langle {u}_{i}(t)Q_{i}(t) \rangle_c - \langle {u}_{i+1}(t)Q_{i}(t) \rangle_c \right]
\end{eqnarray} 
which, however, does not close. It is not difficult to see that the evolution equation for $\langle {u}_{j}(t)Q_{i}(t) \rangle_c$ also does not close and, as a result, it does not seem possible to obtain an exact expression for the current fluctuations. 

To make analytical progress, below we propose a closure scheme that allows us to efficiently handle Eq.~\eqref{Q_sq_time-evolution} and calculate the desired quantities far from criticality quite accurately. On the hydrodynamic scale, when the fluctuations around the global steady-state profile are small, the gradient of the non-conserved  operator ${u}_i$ appearing in the local diffusive current operator ${J}_{i}(t) = (1/2)(u_i - u_{i+1})$ from Eq. \eqref{Q_time-evolution}, is assumed to be governed by the gradient of the conserved density, or local mass operator ${m}_i$, 
\begin{eqnarray}\label{closure_approximation}
 {J}_{i}(t) \simeq D(\overline{\rho}) \left[{m}_{i}(t) -  {m}_{i+1}(t) \right],
\end{eqnarray}
where the bulk-diffusion coefficient $D(\overline{\rho})$ is defined already in Eq.~\eqref{bulk-diffusivity}. \rev{On diffusive scales, the r.h.s. of Eq.~\eqref{closure_approximation} is of $\mathcal{O}(1/L)$ and the error made in this approximation is a subleading term of $\mathcal{O}(1/L^{3/2})$ [see Sec. II of SM].}
As a result, for an arbitrary observable $ {O}(t')$, we can write, \rev{to the leading order},
\begin{eqnarray}\label{closure_1}
    \langle  {J}_{i}(t)  {O}(t') \rangle \simeq D(\overline{\rho}) \left[   \langle  {m}_{i}(t)  {O}(t') \rangle - \langle  {m}_{i+1}(t)  {O}(t') \rangle  \right].
\end{eqnarray}
Under the above (approximate) closure scheme, the \rev{r.h.s.} of Eq.~\eqref{Q_sq_time-evolution} now involves the gradient of the mass-current correlation function $\langle m_{i}(t) Q_{j}(t) \rangle_{c}$. As shown in \rev{Sec.~III} of the SM, the evolution equation for this correlation function requires \rev{the expression of the two-point equal-time mass-mass correlation function} $\langle m_{i}(t) m_{j}(t) \rangle_{c}$, whose evolution equation, however, closes, and can be solved to yield the analytical expression for $\langle m_j(t)Q_i(t) \rangle_c$.
Using these results, we then arrive at the following evolution equation,
\newrev{\begin{eqnarray}\label{time-evolution-Q-Q-general_3}
\frac{d}{dt}\langle Q^{2}(t) \rangle_c &\simeq& 
z({\bar \rho}) - \frac{z({\bar \rho})}{L} \sum_{n=1}^{L-1}(1-e^{-\lambda_{n}D(\overline{\rho})t}).
\end{eqnarray}}
\rev{where $\lambda_{n}=2 (1-\cos q_n)$ with $q_n= 2\pi n/L$.} Notably, we have dropped the site index as the system is translationally invariant. 
Integrating the above equation over time, we finally obtain the variance of the time-integrated bond current as 
\begin{eqnarray}\label{cf}
       \langle Q^{2}(t) \rangle_c &\simeq& \frac{2 \chi({\bar \rho})}{L}\left[t+\sum_{n=1}^{L-1}\frac{1}{\lambda_{n}D(\overline{\rho})}\left(1 - e^{-\lambda_{n} D(\overline{\rho}) t} \right)\right],\nonumber \\ 
\end{eqnarray}
where the mobility $\chi({\bar \rho})={z({\bar \rho})}/{2}$ is simply proportional to the fugacity [as can be seen on comparing Eq.~\eqref{var1} and Eq.~\eqref{asymptotic_Q3} below].


\noindent{\bf Current fluctuations below criticality:} For ${\bar \rho} < {\bar \rho}_c$ where the fugacity is below one and the bulk-diffusion coefficient is finite everywhere, we use the theory developed above to understand the current fluctuations. We are mainly interested in the behavior of $\langle Q^{2}(t) \rangle_c$ on hydrodynamic time scale. But we mention that it scales linearly with time for $t \ll 1/D$, see \rev{Sec.~IV} of the SM for details. 

 \begin{figure}[t]
           \centering
         \includegraphics[width=1\linewidth]{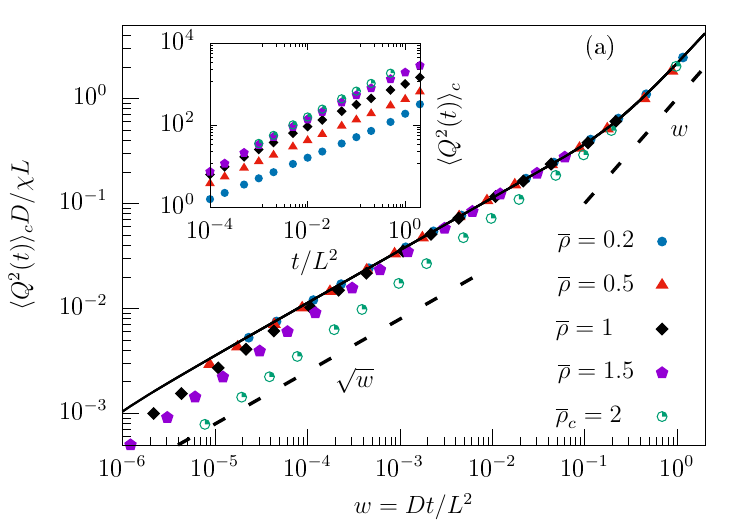}
         \includegraphics[width=1\linewidth]{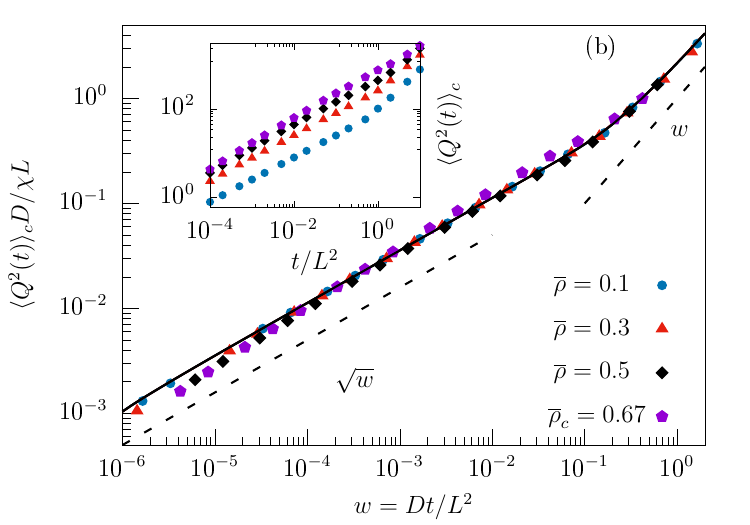}
         \caption{Scaled time-integrated bond-current fluctuation obtained from simulations (points) is compared with Eq.~\eqref{cf_scaling_relation} (line) at various densities for $b=5/2$ (top) and $7/2$ (bottom). The dashed lines represent the short- and long-time behaviors from Eqs.~\eqref{asymptotic_f1} and \eqref{asymptotic_f2}. The inset shows the unscaled time-integrated bond-current fluctuation $\langle Q^{2}(t) \rangle_c$ as a function of $t/L^{2}$ for the same set of densities and $b$ values. In both figures, the system size is $L=1000$.}
          \label{fig:current_fluctuation_scaled}
 \end{figure}

In the hydrodynamic scaling regime where $L \to \infty, D t \to \infty$ such that the ratio $w= Dt/L^{2}$ is finite, Eq.~\eqref{cf} can be written as
\begin{eqnarray}\label{cf_scaling_relation}
\frac{ D({\bar \rho}) \langle Q^{2}(t)\rangle_c}{2 \chi({\bar \rho}) L} = \mathcal{F}\left( \frac{D t}{L^2} \right).
\end{eqnarray}
The scaling function is given by
 \begin{eqnarray}
\label{cf_scaling_fn}
\mathcal{F}(w) &=& \lim_{L \to \infty} \left[ w + \frac{1}{L^{2}}\sum_{n=1}^{L-1} \frac{1}{\lambda_{n}}\left(1- e^{-\lambda_{n} w L^{2}} \right) \right],
\label{scalfn1} \\
&\simeq& w + \sqrt{\frac{w}{\pi}}  \frac{1-e^{-4\pi^2 w}}{4 \pi^2} {\rm{erfc}}(2\pi \sqrt{w}), \label{scafn2}
\end{eqnarray} 
where the last equation is obtained by approximating the sum on the RHS of Eq.~\eqref{scalfn1} by an integral, and ${\rm erfc}(x)$ is the complementary error function. Using the relevant expansion of ${\rm erfc}(x)$ in Eq.~\eqref{scafn2}, we obtain
       \bsn
{\mathcal{F}(w)  \approx} 
\sqrt{\frac{w}{\pi}} &~~~ {\rm for}~~~ $w \ll 1$ , \label{asymptotic_f1} \\
w           &~~~ {\rm for}~~~  $w \gg 1$ .\label{asymptotic_f2} 
\esn  

In other words, the second cumulant of the time-integrated current has the following temporal and system-size scaling,
\bsn
{\langle Q^{2}(t)\rangle_c  \approx} 
\frac{2\chi({\bar \rho})}{\sqrt{D({\bar \rho}) \pi}} \sqrt{t} &~~~ {\rm for}~~~ $1/D \ll t \ll L^{2}/D$ , \label{asymptotic_Q2} \\
\frac{2 \chi({\bar \rho})}{L}t            &~~~ {\rm for}~~~  $t \gg L^{2}/D$ \label{asymptotic_Q3} 
\esn   
which are the same as described in the Introduction at short and long times. Here, in addition to the above scaling, an analytical expression for the amplitude of the temporal growth is also obtained.  \rev{Fig.~(\ref{fig:current_fluctuation_scaled}) shows that the theory developed above agrees very well with the simulation results for ${\bar \rho} < {\bar \rho}_c$.}
Notably, although the ZRP does not have any hardcore constraint, the behavior of the current fluctuation in the ZRP below criticality is still qualitatively similar to that observed in single-file transport \cite{DeMasi:2002, Chakraborty:2024}.

\noindent{\bf Current fluctuations at criticality:} \rev{At the critical point, as the fugacity or, equivalently, the mobility remains finite for all $b>2$, from the above discussion, we expect that the long-time behavior in Eq.~\eqref{asymptotic_Q3} continues to hold.} However, as the bulk-diffusion coefficient vanishes for $2 < b < 3$ in the thermodynamic limit, due to Eq.~\eqref{asymptotic_Q2}, we anticipate a change in the scaling at short times as explained below. 

\begin{figure}[t]
           \centering
           \includegraphics[width=1\linewidth]{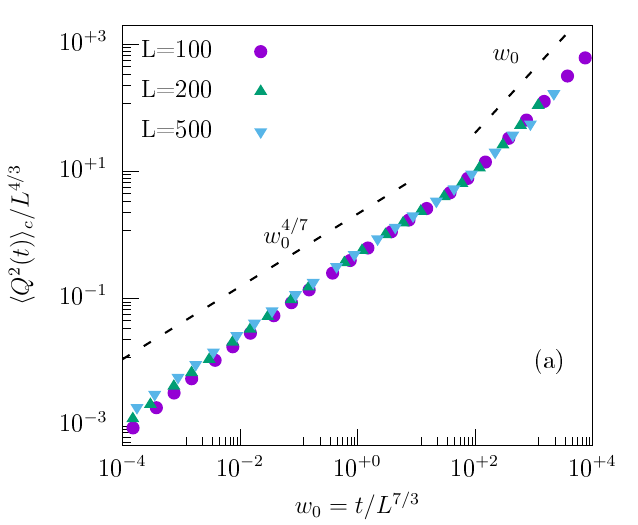}
          \includegraphics[width=1\linewidth]{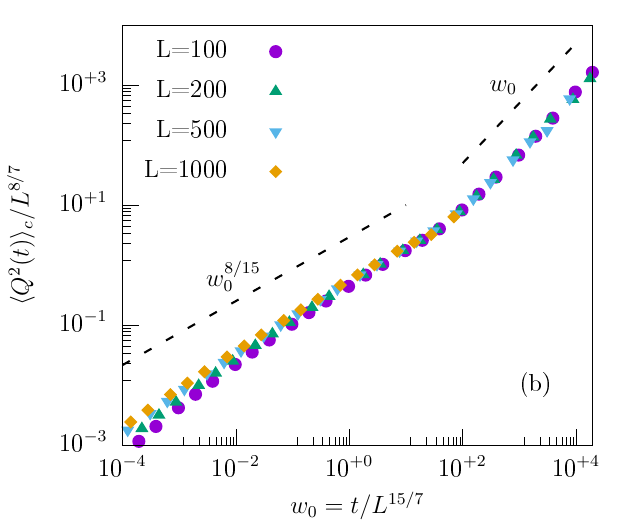}
         \caption{Scaled second cumulant of current as a function of the scaled time for various system sizes at the critical density ${\bar \rho}={\bar \rho}_c$ for $b=5/2$ (top panel) and $11/4$ (bottom panel), where the points show simulation data and the dashed lines are the predictions from Eqs.~\eqref{critI1} and \eqref{critI2}.}
          \label{fig:current_fluctuation_scaled_criticality}
 \end{figure}
\rev{We begin by characterizing how the bulk-diffusion coefficient $D(\bar\rho)$ scales with system size $L$ for $2<b<3$ at the critical point $\bar{\rho} = \bar{\rho}_{c}$. For this purpose, we adopt the following scaling ansatz for the stationary single-site mass distribution function in the near-critical regime:
\begin{eqnarray}\label{scaling_ansatz_pm}
    p(m; \bar\rho,L)= m^{-b} F\left( \frac{m}{L^\gamma}, m (\bar\rho_c-\bar\rho)^\delta \right)
\end{eqnarray}
where the factor $m^{-b}$ takes care of the distribution at criticality in an infinitely large system, and $F(0,y)$ and $F(x,0)$ are, respectively, the scaling function in the thermodynamic limit in the fluid phase and at criticality in a finite system, which have been obtained in \cite{Evans:2006}. From their Eqs.~(29), (30), and (69), we find that $\delta = (b-2)^{-1}$ and $\gamma = (b-1)^{-1}$ for $2 < b < 3$, while for $b \geq 3$, $\delta = 1$ and $\gamma = 1/2$. Eq.~\eqref{scaling_ansatz_pm} along with these results imply the following scaling behavior:
\bsn
{\bar\rho_c -\bar\rho \sim }
   L^{-\frac{b-2}{b-1}}   & {\rm for}~ $2 < b < 3,$ \label{rho_L_scaling_1} \\
  L^{-1/2}           & {\rm for}~  $b \geq 3.$ \label{rho_L_scaling_2}
\esn  
Now using the previously derived near-critical density-fugacity relation from \cite{Priyanka:2014}, obtained by expanding Eq.~\eqref{fugacity} about $\bar \rho = \bar \rho_c$ and $z=1$, and the definition of $D(\bar\rho)$ in Eq.~\eqref{bulk-diffusivity}, we find that as the system approaches criticality, $D(\bar\rho)$ vanishes as $(\bar{\rho}_c - \bar{\rho})^{\frac{3-b}{b-2}}$ for $2 < b < 3$, while it remains constant for $b > 3$ (see Sec. I of the SM). The above observation, along with Eq.~\eqref{rho_L_scaling_1}, immediately yields that the bulk-diffusion coefficient exhibits the following scaling at the critical point,
\bsn
{D(\bar \rho_c) \equiv D_{L} \sim }
   L^{-\frac{3-b}{b-1}}   & {\rm for}~ $2 < b < 3,$ \label{D_L_scaling_1} \\
  \mathcal{O}(1)           & {\rm for}~  $b \geq 3.$ \label{D_L_scaling_2}
\esn}

Assuming a scaling hypothesis  akin to Eq.~\eqref{cf_scaling_relation}, albeit now at the criticality, and the above finite-size scaling of bulk-diffusion coefficient, we have the following scaling form for  $\langle Q^2(t) \rangle_c$ for densities at criticality: 
 \begin{equation}
 \langle Q^2(t) \rangle_c=\frac{L}{D_L} {\cal F}_c\left(\frac{D_L t}{L^{2}} \right)=L^\beta {\cal F}_c\left(\frac{t}{L^{1+\beta}} \right).
\end{equation}
Here ${\cal F}_c(w) \sim w$ for $w \gg 1$ as the long-time behavior is not affected by the vanishing bulk-diffusion coefficient and is solely determined by the mobility. Also, we have ${\cal F}_c(w) \sim w^{\frac{\beta}{1+\beta}}$ for $w \ll 1$ as the short-time behavior, which corresponds to taking the thermodynamic limit, must be independent of $L$. \rev{Then, by using $L/D_L \sim L^\beta$, at the critical point, we have $\beta=1$ for $b \geq 3$ as the bulk-diffusion coefficient is of order unity, while we obtain $\beta=2/(b-1)$ for $2 < b < 3$ on using Eq.~\eqref{D_L_scaling_1} for the $L$-dependence of the bulk-diffusion coefficient close to the critical point.}

The bottom panel of Fig.~\ref{fig:current_fluctuation_scaled} for $b=7/2$ shows that, at the critical density, the scaling behavior in Eq.~\eqref{cf_scaling_relation} continues to hold reasonably well, but the top panel for $b=5/2$ shows that the variance of the time-integrated current does not follow Eq.~\eqref{cf_scaling_relation}. \rev{However, the data in Fig.~\ref{fig:current_fluctuation_scaled_criticality} for $2 < b < 3$ and at the critical density provides strong support for the following scaling: 
\bsn
{\langle Q^{2}(t)\rangle_c=L^{\frac{2}{b-1}} {\cal F}_c\left( \frac{t}{L^{\frac{b+1}{b-1}}}\right)  \propto }
  t^{\frac{2}{b+1}}   & {\rm for}~ $t \ll L^{\frac{b+1}{b-1}}$ \label{critI1} \\
  \frac{t}{L}            & {\rm for}~  $t \gg L^{\frac{b+1}{b-1}}$ \label{critI2}
\esn
The above expression} shows that at short times, the variance of the cumulative current grows faster than the generic $\sqrt{t}$ scaling away from the critical point, and the crossover to the linear scaling occurs at a time scale larger than the diffusive scale. Furthermore, the growth exponent at short times varies continuously with $b$ for $2 < b < 3$, and we recover the $\sqrt{t}$ scaling for $b \geq 3$. 
\rev{Indeed, the anomalous growth law as in Eq.~\eqref{critI1} implies violation of the usual hydrodynamic description at criticality for $ 2 < b < 3 $. 
On physical grounds, this can be explained by the behavior of mass fluctuations, which, as suggested by Eq.~\eqref{scaling_ansatz_pm} and the subsequent discussions, diverge in the range $ 2 < b < 3 $ while remaining finite for $b>3$. Notably, Eq.~\eqref{bulk-diffusivity} suggests the bulk-diffusion coefficient is inversely related to the variance $\sigma^{2}(\bar\rho)$ of onsite mass  through a fluctuation relation $D(\bar\rho)={z(\bar\rho)}/{\sigma^{2}(\bar\rho)}$, where  $\sigma^{2}(\bar\rho)=\langle m^2 \rangle_c=z(d\bar\rho/dz)$. As a result, the diverging mass fluctuation essentially leads to the bulk-diffusion coefficient vanishing in the range $2 < b < 3$, as quantified in Eq.~\eqref{D_L_scaling_1}. }


\noindent{\bf Summary and concluding remarks:} In this study,  we have shown that the steady-state current fluctuations show anomalous scaling at the critical point when the parameter $2 < b < 3$, and the origin of this result lies in the vanishing bulk-diffusion coefficient in the thermodynamic limit. Such a behavior of the bulk-diffusion coefficient has also been observed in a ZRP with sitewise quenched disorder where it remains finite or vanishes depending on the distribution of hopping rates \cite{Barma:2002}; indeed it would be quite interesting to study the dynamics of current fluctuations in such disordered systems. 

\rev{The results here are obtained using a dynamical theory where we} introduced a closure (approximate) scheme and they are supported through simulations. 
{We emphasize that, although we have considered here only the ZRP, our results are expected to have a broader validity for the systems at criticality, where the bulk-diffusion coefficient vanishes (as expected in a clustering transition), but the mobility remains finite.
Indeed, it would be of great interest if the scaling function for the current fluctuations at the critical point can be calculated using a microscopic theory.}
 Here, our analysis is limited to the second cumulant of the time-integrated current, and extending these results to the full distribution of the time-integrated current, or higher-order statistics, using a detailed hydrodynamic theory is also certainly desirable.


\noindent{\bf Acknowledgements:} { We thank Gunter Sch\"utz and Tridib Sadhu for useful discussions.} TC and KJ  thank S. N. Bose National Centre for Basic Sciences (SNBNCBS), Kolkata for hospitality, and PP thanks SNBNCBS, Kolkata for funding during the conference Steady State Phenomena in Soft Matter, Active and Biological Systems (2023), where this work was initiated.
We also thank the International Centre for Theoretical Sciences (ICTS), Bengaluru for hospitality during the 9th Indian Statistical Physics Community Meeting (ICTS/ISPCM2024/4) that facilitated our discussions.

 \bibliography{zrp}

 \end{document}


\newcommand{\rar}{\rightarrow}
\newcommand{\lar}{\leftarrow}
\newcommand{\rlh}{\rightleftharpoons}
\newcommand{\eref}[1]{Eq.~(\ref{#1})}%
\newcommand{\Eref}[1]{Equation~(\ref{#1})}%
\newcommand{\fref}[1]{Fig.~\ref{#1}} %
\newcommand{\Fref}[1]{Figure~\ref{#1}}%
\newcommand{\sref}[1]{Sec.~\ref{#1}}%
\newcommand{\Sref}[1]{Section~\ref{#1}}%
\newcommand{\aref}[1]{Appendix~\ref{#1}}%

\renewcommand{\ni}{{\noindent}}
\newcommand{\dprime}{{\prime\prime}}
\newcommand{\be}{\begin{equation}}
\newcommand{\ee}{\end{equation}}
\newcommand{\bea}{\begin{eqnarray}}
\newcommand{\eea}{\end{eqnarray}}
\newcommand{\nn}{\nonumber}
\newcommand{\bk}{{\bf k}}
\newcommand{\bQ}{{\bf Q}}
\newcommand{\q}{{\bf q}}
\newcommand{\s}{{\bf s}}
\newcommand{\bN}{{\bf \nabla}}
\newcommand{\bA}{{\bf A}}
\newcommand{\bE}{{\bf E}}
\newcommand{\bj}{{\bf j}}
\newcommand{\bJ}{{\bf J}}
\newcommand{\bs}{{\bf v}_s}
\newcommand{\bn}{{\bf v}_n}
\newcommand{\bv}{{\bf v}}
\newcommand{\la}{\left\langle}
\newcommand{\ra}{\right\rangle}
\newcommand{\dg}{\dagger}
\newcommand{\br}{{\bf{r}}}
\newcommand{\brp}{{\bf{r}^\prime}}
\newcommand{\bq}{{\bf{q}}}
\newcommand{\hx}{\hat{\bf x}}
\newcommand{\hy}{\hat{\bf y}}
\newcommand{\bS}{{\bf S}}
\newcommand{\cU}{{\cal U}}
\newcommand{\cD}{{\cal D}}
\newcommand{\bR}{{\bf R}}
\newcommand{\pll}{\parallel}
\newcommand{\sumr}{\sum_{\vr}}
\newcommand{\cP}{{\cal P}}
\newcommand{\cQ}{{\cal Q}}
\newcommand{\cS}{{\cal S}}
\newcommand{\ua}{\uparrow}
\newcommand{\da}{\downarrow}
\newcommand{\red}{\textcolor {red}}
\newcommand{\blue}{\textcolor {blue}}
\newcommand{\1}{{\oldstylenums{1}}}
\newcommand{\2}{{\oldstylenums{2}}}
\newcommand{\mDelta}{\varepsilon}
\renewcommand{\thepage}{S-\arabic{page}}
\newcommand{\m}{\tilde m}
\def\lsim {\protect \raisebox{-0.75ex}[-1.5ex]{$\;\stackrel{<}{\sim}\;$}}
\def\gsim {\protect \raisebox{-0.75ex}[-1.5ex]{$\;\stackrel{>}{\sim}\;$}}
\def\lsimeq {\protect \raisebox{-0.75ex}[-1.5ex]{$\;\stackrel{<}{\simeq}\;$}}
\def\gsimeq {\protect \raisebox{-0.75ex}[-1.5ex]{$\;\stackrel{>}{\simeq}\;$}}

\title{\large{Supplemental Material: \\Current fluctuations in symmetric zero-range process below and at critical density}}
\author{Tanmoy Chakraborty\textsuperscript{$\star$}, Punyabrata Pradhan\textsuperscript{$\star$} and Kavita Jain\textsuperscript{$\dagger$}} 
\affiliation{
${}^\star$ Department of Physics of Complex Systems, S. N. Bose
   National Centre for Basic Sciences, Block-JD, Sector-III, Salt Lake,
   Kolkata 700106, India
\\
${}^\dagger$ Theoretical Sciences Unit,
Jawaharlal Nehru Centre for Advanced Scientific Research,
Bangalore 560064, India}

\maketitle
\nopagebreak

\section{Macroscopic transport coefficients}
\label{SM_Dchi}

\begin{figure}[H]
           \centering
         \includegraphics[width=0.5\linewidth]{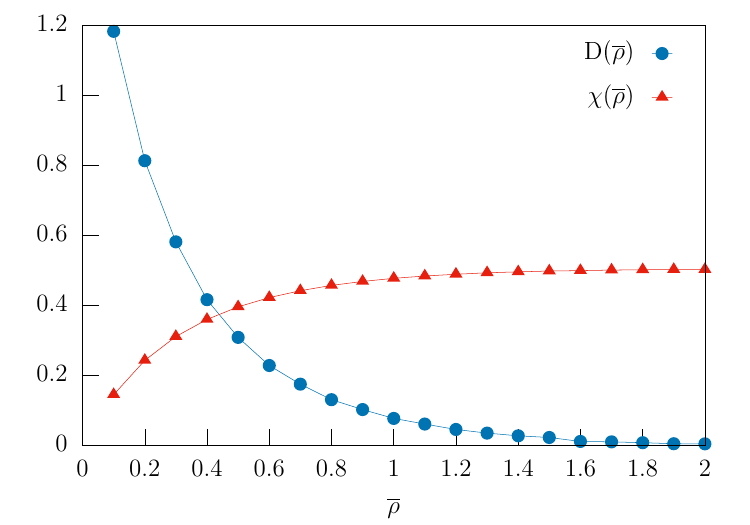}
        \caption{Figure shows that with increasing density, the bulk-diffusion constant decreases and collective particle mobility increases towards one half; here $b=5/2$.}
          \label{fig:transport_coefficients}
 \end{figure}

 
On expanding the l.h.s. and r.h.s. of Eq.~(6) of the main text about the critical density and fugacity $z=1$, respectively, we find that, to leading order in $1-z$, 
  \bsn
 { \bar\rho-\bar\rho_c=}
 - \pi \text{\textcolor{black}{cosec}}(b \pi) (b-1)^2 (1-z)^{b-2}~,~2 < b < 3 \label{rhoz1}\\
 8 (1-z) \ln (1-z)~,~ b=3 \\
  \frac{-(b-1)^2 (1-z)}{(b-3) (b-2)^2}~,~ b > 3 \label{rhoz3}
  \esn
Using the definition of bulk-diffusion constant given in Eq.~(11) and the above equation, we  immediately obtain
\begin{center}
\bsn
{D=\frac{dz}{d \bar\rho}=}
\frac{(1-z)^{3-b}}{ \pi \text{\textcolor{black}{cosec}}(b \pi) (b-1)^2 (b-3)} \textcolor{black}{\sim (\bar\rho_c - \bar\rho)^{\frac{3-b}{b-2}}}  ~,~2 < b < 3\\
\frac{1}{8 |\ln (1-z)|} ~,~b=3 \\
\frac{(b-3) (b-2)^2}{(b-1)^2 }~,~ b > 3
\esn  
\end{center}
which shows that the bulk-diffusion constant vanishes for $2 < b \leq 3$ at the critical density, as also indicated in  
Fig.~\ref{fig:transport_coefficients}.

 \section{Statistics of fluctuating current}

We use the following decomposition of the instantaneous current:
\begin{equation}
\mathcal{J}_{r}(t)= \frac{dQ_{r}(t)}{dt} = J_{r}(t) + J^{(fl)}_{r}(t),
\end{equation}
where $J_{r}(t)$ is the ``hydrodynamic'' contribution, as in Eq.~(9) of the main text, and $J^{(fl)}_{r}(t)$ being the ``noise'' or fluctuating component. One can immediately see that
\begin{eqnarray}\label{fluc_average}
    \langle J^{(fl)} \rangle =0.
\end{eqnarray}
Moreover, using the above decomposition, the space-time correlation $\mathcal{C}^{J^{fl}J^{fl}}_{r}(t, 0)= \langle J^{fl}_{r}(t) J^{fl}_{0}(0) \rangle_c$ can be exactly calculated within the approximation scheme as in Eq.~(14) of the main text. 
Performing further algebraic manipulations, we obtain the following:
\begin{eqnarray}
\mathcal{C}^{\mathcal{J}\mathcal{J}}_{r}(t,t')= z(\bar\rho) \delta_{r,0} \delta(t-t') + \mathcal{C}^{J\mathcal{J}}_{r}(t,t') \Theta(t-t') + \mathcal{C}^{\mathcal{J} J}_{r}(t,t') \Theta(t'-t).\label{eq_3}
\end{eqnarray}
Clearly the above solution for $t \geq t'=0$ is given by 
\begin{eqnarray}
\mathcal{C}^{\mathcal{J}\mathcal{J}}_{r}(t,0)=z(\bar\rho) \delta_{r,0} \delta(t) + \mathcal{C}^{J\mathcal{J}}_{r}(t,0),\label{eq_4}
\end{eqnarray}
leading to, albeit within the approximation Eq. (14) of main text,
\begin{eqnarray}
    \mathcal{C}^{J^{fl}J^{fl}}_{r}(t, 0) = z(\bar\rho) \delta_{r,0} \delta(t),
\end{eqnarray}
which exactly captures the fluctuation-dissipation theorem (FDT) valid for equilibrium
systems like the symmetric ZRP.
From the above equation along with Eq.~\eqref{fluc_average}, we can see that the fluctuating component $J^{(fl)}$ exhibits a Gaussian white-noise-like behavior with zero mean and variance given by $z(\bar\rho)$. On large spatiotemporal scales (diffusive scaling limit discussed below), the diffusive current operator $J_r \equiv (1/2)(u_r - u_{r+1})$ should be equivalent to the ``approximate'' current operator $J_r^{approx} \equiv D(\bar \rho) (m_r - m_{r+1})$ - a microscopic version of the Fick's law as approximated in Eq. (14) of the main text. 
In the diffusive scaling limit  $r \to x=r/L$ and $t \to \tau=t/L^2$, currents (`gradients') will scale as follows:
\begin{equation}
    J_r(t) = \frac{1}{L} j(x,\tau)
\end{equation}
and
\begin{equation}
   J_r^{approx}(t) = \frac{1}{L} j^{approx}(x,\tau); 
\end{equation}
 the noise current, on the other hand, transforms as 
 \begin{equation}
J^{fl}_r(t) = \frac{1}{L^{3/2}} j^{fl}(x,\tau)     
 \end{equation}
due to the short-ranged (delta-correlated) correlations.
Essentially, on a statistical level and far from criticality, while the diffusive currents $J_r \sim {\cal O}(1/L)$ and $J^{approx}_r \sim {\cal O}(1/L)$, the difference $J_r - J^{approx}_r \sim {\cal O}(1/L^{3/2})$, however, is much smaller and could be ignored in the diffusive scaling limit.


\section{Calculation of time-integrated bond current cumulant $\langle Q^{2}(t) \rangle_c$}

Upon implementing the closure scheme described in Eq.~(15) of the main text, the exact time-evolution equation for $\langle Q^{2}(t) \rangle_c$ in Eq.~(13) of the main text reads as follows:
\begin{eqnarray}\label{time-evolution-Q-Q-general_2}
\frac{d}{dt}\langle Q^{2}_{i}(t) \rangle_c \simeq \frac{1}{2}\left[ \langle u_{i} \rangle + \langle u_{i+1} \rangle \right] + 2 D(\overline{\rho}) \left[ \langle m_{i}(t) Q_{i}(t) \rangle_c -  \langle m_{i+1}(t) Q_{i}(t) \rangle_c \right].
\end{eqnarray}
Therefore, calculating $\langle Q^{2}_{i}(t) \rangle_c$ essentially reduces to computing the equal-time mass-current correlation function $\langle m_j(t)Q_i(t) \rangle_c = \langle m_j(t)Q_i(t) \rangle - \langle m_j(t)\rangle \langle Q_i(t) \rangle$, which we will now proceed to calculate.

\subsection{Equal time mass-current correlation function}
\label{SM_mJ}

Below, we list all the possible combinations that result in the update of the product $m_{j}(t)Q_{i}(t)$ within the infinitesimal time window $[t, t+dt]$, as given by

\begin{eqnarray}\label{m_Q_update_eqn}
m_{j}(t+dt)Q_{i}(t+dt) = 
\left\{
\begin{array}{ll}
(m_{j}(t)-1)(Q_{i}(t) + 1) \vspace{0.25 cm}, ~~ & {\rm prob.}~~ \frac{1}{2}	u_{i} \delta_{i,j} dt, \\
(m_{j}(t)-1)(Q_{i}(t) - 1) \vspace{0.25 cm}, ~~ & {\rm prob.}~~ \frac{1}{2}	u_{i+1} \delta_{i,j-1} dt, \\
(m_{j}(t)+1)(Q_{i}(t) - 1) \vspace{0.25 cm}, ~~ & {\rm prob.}~~ \frac{1}{2}u_{i+1} \delta_{i,j} dt , \\
(m_{j}(t)+1)(Q_{i}(t) + 1) \vspace{0.25 cm}, ~~ & {\rm prob.}~~ \frac{1}{2}u_{i} \delta_{i,j-1} dt , \\
(m_{j}(t)-1)Q_{i}(t)  \vspace{0.25 cm}, ~~ & {\rm prob.}~~ \frac{1}{2}u_{j}(2-\delta_{i,j}-\delta_{i,j-1}) dt \\
(m_{j}(t)+1)Q_{i}(t)  \vspace{0.25 cm}, ~~ & {\rm prob.}~~ \frac{1}{2} \Big\{u_{j+1} (1-\delta_{i,j}) + u_{j-1} (1-\delta_{i,j-1}) \Big\} dt. \\
m_{j}(t)(Q_{i}(t) + 1) \vspace{0.25 cm}, ~~ & {\rm prob.}~~ \frac{1}{2}u_{i}(1-\delta_{i,j}-\delta_{i,j-1}) dt, \\
m_{j}(t)(Q_{i}(t) - 1) \vspace{0.25 cm}, ~~ & {\rm prob.}~~ \frac{1}{2}u_{i+1}(1-\delta_{i,j}-\delta_{i,j-1}) dt , \\
m_{j}(t)Q_{i}(t) \vspace{0.25 cm}, ~~ & {\rm prob.}~~ 1- \Sigma dt , \\
\end{array}
\right.
\end{eqnarray}
where $\Sigma$ represents the sum of exit rates and is given by,
\begin{eqnarray}
\Sigma = \frac{1}{2}(u_{j+1} + u_{j-1} + 2 u_{j}) + \frac{1}{2} (u_{i+1} + u_{i}) (1-\delta_{i,j}-\delta_{i,j-1}).
\end{eqnarray}
Using the above update rule in Eq.~\eqref{m_Q_update_eqn}, the time-evolution equation of $\langle	m_{j}(t)Q_{i}(t)\rangle$ is calculated to be,
\begin{eqnarray}\label{m_Q_evolution_1}
\frac{d}{dt}\langle m_{j}(t)Q_{i}(t) \rangle = \frac{1}{2}\left[ \langle u_{i} \rangle + \langle u_{i+1} \rangle \right] (\delta_{i,j-1}-\delta_{i,j})+ \left[ \langle Q_{i}(t)J_{j-1}(t) \rangle - \langle Q_{i}(t)J_{j}(t) \rangle \right]+\langle m_{j}(t)J_{i}(t) \rangle.
\end{eqnarray}
where, 
\begin{eqnarray}
J_i(t)=\frac{1}{2} [u_i(t)-u_{i+1}(t)]
\end{eqnarray}
Moreover, by applying the continuity equation, the time-evolution of the local mass $m_{j}(t)$ can be written as,
\begin{eqnarray}\label{continuity_mass}
\frac{d\langle m_{j} \rangle}{dt} = \langle J_{j-1}(t) \rangle - \langle J_{j}(t) \rangle.
\end{eqnarray} 
Finally, by using Eqs.~\eqref{m_Q_evolution_1} and \eqref{continuity_mass}, we obtain the time-evolution of $\langle	m_{j}(t)Q_{i}(t)\rangle_{c}$ in the following manner:
\begin{eqnarray}
\frac{d}{dt}\langle m_{j}(t)Q_{i}(t) \rangle_c &=& \frac{d}{dt}\langle m_{j}(t)Q_{i}(t) \rangle - \left[ \frac{d \langle m_{j}\rangle}{dt}  \left\langle Q_{i}(t) \right\rangle + \langle m_{j}(t) \rangle \frac{d \langle Q_{i} \rangle}{dt}  \right], \\ &=& \frac{1}{2}\left[ \langle u_{i} \rangle + \langle u_{i+1} \rangle \right] (\delta_{i,j-1}-\delta_{i,j})+ \left[ \langle Q_{i}(t) J_{j-1}(t) \rangle_c - \langle Q_{i}(t)J_{j}(t) \rangle_c \right]+\langle m_{j}(t)J_{i}(t) \rangle_c,
\label{m_Q_evolution_3} \\ &=& D(\overline{\rho}) \Delta^{2}_{j} \langle m_{j}(t)Q_{i}(t) \rangle_{c} + D(\overline{\rho}) \Delta_{j} \langle m_{j}(t)m_{i}(t) \rangle_{c} - \frac{1}{2}\left[ \langle u_{i} \rangle + \langle u_{i+1} \rangle \right] \Delta_{j}\delta_{i,j}.\label{m_Q_evolution_2}
\end{eqnarray}  
In the above equation, $\Delta_{j}$ and $\Delta^{2}_{j}$ represent the discrete gradient and Laplacian operators, respectively, and they are defined as follows:
\begin{eqnarray}
\Delta_{j}f_{j}&=&f_{j}-f_{j-1},\\
\Delta^{2}_{j}f_{j}&=&f_{j+1}+f_{j-1}-2f_{j}.
\end{eqnarray}
Clearly, Eq.~\eqref{m_Q_evolution_2} indicates that in order to solve for the mass-current correlation function, we must first determine the steady-state mass-mass correlation function $\langle m_{j}(t)m_{i}(t) \rangle_{c} = \langle m_{j}(t)m_{i}(t) \rangle - \langle m_{j}(t) \rangle \langle m_{i}(t) \rangle$, which is addressed in the following section.

 
\subsection{Equal-time mass-mass correlation function}
\label{SM_mm}

We begin by writing down the events that corresponds to the update of the product $m_{j}(t)m_{i}(t)$ in the infinitesimal time interval $[t, t+dt]$, as given by
\begin{eqnarray}\label{m_m_update_eqn}
m_{j}(t+dt)m_{i}(t+dt) = 
\left\{
\begin{array}{ll}
(m_{j}(t)-1)^{2} \vspace{0.25 cm}, ~~ & {\rm prob.}~~ 	u_{j} \delta_{i,j} dt, \\
(m_{j}(t)+1)^{2} \vspace{0.25 cm}, ~~ & {\rm prob.}~~ \frac{1}{2}	(u_{j+1} + u_{j-1}) \delta_{i,j} dt, \\
(m_{j}(t)+1)(m_{i}(t) - 1) \vspace{0.25 cm}, ~~ & {\rm prob.}~~ \frac{1}{2}(u_{j+1} \delta_{i,j+1} + u_{j-1} \delta_{i,j-1}) dt , \\
(m_{j}(t)-1)(m_{i}(t) + 1) \vspace{0.25 cm}, ~~ & {\rm prob.}~~ \frac{1}{2}(u_{i+1} \delta_{j,i+1} + u_{i-1} \delta_{j,i-1}) dt , \\
(m_{j}(t)-1)m_{i}(t)  \vspace{0.25 cm}, ~~ & {\rm prob.}~~ \frac{1}{2}u_{j}(2-2\delta_{i,j}-\delta_{i,j-1}-\delta_{i,j+1}) dt \\
(m_{j}(t)+1)m_{i}(t)  \vspace{0.25 cm}, ~~ & {\rm prob.}~~ \frac{1}{2} \Big\{u_{j+1} (1-\delta_{i,j}-\delta_{i,j+1}) + u_{j-1} (1-\delta_{i,j}-\delta_{i,j-1}) \Big\} dt. \\
m_{j}(t)(m_{i}(t) + 1) \vspace{0.25 cm}, ~~ & {\rm prob.}~~ \frac{1}{2}\Big\{u_{i+1} (1-\delta_{i,j}-\delta_{j,i+1}) + u_{i-1} (1-\delta_{j,i}-\delta_{j,i-1}) \Big\} dt, \\
m_{j}(t)(m_{i}(t) - 1) \vspace{0.25 cm}, ~~ & {\rm prob.}~~ \frac{1}{2}u_{i}(2-2\delta_{i,j}-\delta_{j,i-1}-\delta_{j,i+1}) dt , \\
m_{j}(t)m_{i}(t) \vspace{0.25 cm}, ~~ & {\rm prob.}~~ 1- \Sigma dt . \\
\end{array}
\right.
\end{eqnarray}
Here, the sum of exit rates, i.e., $\Sigma$ is given by,
\begin{eqnarray}
\Sigma= \frac{1}{2}(u_{j+1} + u_{j-1} + 2u_{j}) + \frac{1}{2}(u_{i+1} + u_{i-1} + 2u_{i}) + \frac{1}{2} u_{j+1}(\delta_{i,j+1} - \delta_{i,j}) + \frac{1}{2} u_{j-1}(\delta_{i,j-1} - \delta_{i,j}).
\end{eqnarray}
Using the update rule in Eq.~\eqref{m_m_update_eqn}, the corresponding time-evolution equation for $\langle m_{j}(t)m_{i}(t)\rangle$ is calculated to be,
\begin{eqnarray}\label{m_m_evolution_0}
\frac{d}{dt}\langle m_{i}(t) m_{j}(t) \rangle = \frac{1}{2}\left[ \langle u_{i} \rangle + \langle u_{i+1} \rangle \right] \Delta^{2}_{j} \delta_{i,j} - \Delta_{j}\langle m_{i}(t) J^{(D)}_{j-1}(t) \rangle - \Delta_{i}\langle m_{j}(t) J^{(D)}_{i-1}(t) \rangle.
\end{eqnarray}
Now, by using the above equation along with Eq.~\eqref{continuity_mass}, the time-evolution of equal-time mass-mass correlation function is obtained to satisfy the following equation:
\begin{eqnarray}\label{m_m_evolution_1}
\frac{d}{dt}\langle m_{i}(t) m_{j}(t) \rangle_c = \frac{1}{2}\left[ \langle u_{i} \rangle + \langle u_{i+1} \rangle \right] \Delta^{2}_{j} \delta_{i,j} - \Delta_{j}\langle m_{i}(t) J_{j-1}(t) \rangle_c - \Delta_{i}\langle m_{j}(t) J_{i-1}(t) \rangle_c.
\end{eqnarray}
Notably, upon using the closure approximation, the above equation can be rewritten as,
\begin{eqnarray}\label{m_m_evolution_2}
\frac{d}{dt}\langle m_{i}(t) m_{j}(t) \rangle_c = \frac{1}{2}\left[ \langle u_{i} \rangle + \langle u_{i+1} \rangle \right]  \Delta^{2}_{j} \delta_{i,j} - 2 D(\overline{\rho}) \Delta^{2}_{j}\langle m_{i}(t) m_{j}(t) \rangle_c.
\end{eqnarray}
One can now solve Eq.~\eqref{m_m_evolution_2} to obtain the time-dependent solution of $\langle m_{i}(t) m_{j}(t) \rangle_c$. 

Since, in this work, we want to achieve current fluctuation at the steady state, we need to calculate the steady state value of $\langle m_{i}(t) m_{j}(t) \rangle_c$. To this end, we immediately obtain the steady-state condition by dropping the explicit time dependence from Eq.~\eqref{m_m_evolution_2} and assuming spatial homogeneity $\langle u_{i+1} \rangle = \langle u_{i} \rangle = z(\overline{\rho})$. Consequently, the desired steady-state solution then takes the following form:
\begin{eqnarray}\label{m_m_evolution_3}
 \langle m_{i} m_{j} \rangle_c = \frac{z(\overline{\rho})}{2 D(\overline{\rho})} \delta_{i,j}.
\end{eqnarray}
Note that, the above solution for $\langle m_{i} m_{j} \rangle_c$ immediately allows for calculating the time-evolution equation  of the equal-time mass-current correlation function at the steady state, which can be written as,
\begin{eqnarray}\label{m_Q_evolution_3}
\frac{d}{dt}\langle m_{j}(t)Q_{i}(t) \rangle_{c} = D(\overline{\rho}) \Delta^{2}_{j} \langle m_{j}(t)Q_{i}(t) \rangle_{c}  - \frac{z(\overline{\rho})}{2}\Delta_{j}\delta_{i,j}.
\end{eqnarray}
It is simple to solve Eq.~\eqref{m_Q_evolution_3} in the Fourier basis. To implement this strategy, we first obtain the solution in the Fourier basis, and then perform the inverse Fourier transformation to obtain the solution for $\langle m_j(t)Q_i(t) \rangle$ in real space, which is calculated to be,
\textcolor{black}{\begin{eqnarray}
\langle m_j(t)Q_i(t) \rangle_c = -\frac{z(\overline{\rho})}{2D(\overline{\rho})L}\sum_{n=1}^{L-1}\frac{e^{-iq_{n}r}}{\lambda_{{n}}}(1-e^{iq_{n}})(1-e^{-\lambda_{n} D t}), \label{mQresult}
\end{eqnarray}}
where $q_n=2\pi n/L$ with $n=1,$ $2, \dots, L-1$, $r=j-i$ and $\lambda_{n}$ is the eigenvalue of the discrete Laplacian operator, which is given by
\textcolor{black}{\begin{eqnarray}
\lambda_{n}=2(1-\cos q_n).
\end{eqnarray}}

Finally, plugging in the above steady state solution of equal-time mass-current correlation function of Eq.~\eqref{mQresult} and incorporating the condition of spatial homogeneity $\langle u_{i+1} \rangle = \langle u_{i} \rangle = z(\overline{\rho})$ in Eq.~\eqref{time-evolution-Q-Q-general_2}, we find the steady-state time-integrated current fluctuation to satisfy the following time evolution equation:
\textcolor{black}{\begin{eqnarray}
\frac{d}{dt}\langle Q^{2}_{i}(t) \rangle_c &\simeq&  z(\overline{\rho}) - \frac{z(\overline{\rho})}{L} \sum_{n=1}^{L-1}\frac{(1-e^{-iq_n})}{\lambda_{n}}(1-e^{iq_n})(1-e^{-\lambda_{n} D t})\\ &=& z(\overline{\rho}) -  \frac{z(\overline{\rho})}{L} \sum_{n=1}^{L-1}(1-e^{-\lambda_{n} D t}).\label{time-evolution-Q-Q-general_final_sm}
\end{eqnarray}}
which is Eq.~(16) of the main text. 

\section{Short time behavior of current fluctuations}
\label{shorttime}

\begin{figure}[H]
           \centering
                \includegraphics[width=0.49\linewidth]{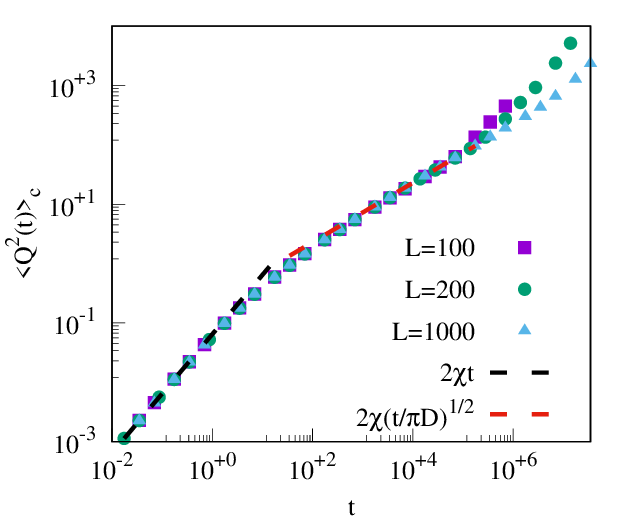}
         \caption{Time-integrated bond-current fluctuations $\langle Q_{i}^{2}(t)\rangle$ against the observation time $t$ for various system sizes $L$ for $\rho=0.5$ and $b=2.5$. The dotted lines demonstrate $L$-independent growth at small times (black dotted line), as defined in Eq.~\eqref{small_time_growth}, and at intermediate regime (red dotted line), as given in Eq.~(22a) of the main text.}
          \label{fig:shorttime}
 \end{figure}

For $Dt \ll 1$, by linearly expanding the exponential term in Eq.~(17) of the main text, we find that
\begin{eqnarray}\label{small_time_growth}
    \lim_{Dt \ll 1} \langle Q_{i}^{2}(t) \rangle \simeq 2\chi({\bar \rho}) t, \label{asymptotic_Q1}
\end{eqnarray}
which  agrees quite well with the numerical data shown in Fig.~\ref{fig:shorttime}.

%
%
%
